\documentclass[aps,prx,twocolumn,final,letterpaper,floatfix]{revtex4-2}

\usepackage{appendix}
\usepackage{graphicx}   
\usepackage{import}
\usepackage{comment}
\usepackage{epstopdf}
\usepackage{bm}
\usepackage{amssymb}
\usepackage{quotes}
\usepackage{braket}
\usepackage{transparent}
\usepackage{dcolumn}
\usepackage{multirow}
\usepackage{cancel} 
\usepackage{mdframed}
\usepackage{color}
\usepackage{bm}
\usepackage{dsfont}
\usepackage{slashed}
\usepackage{enumitem}
\usepackage{amsmath} 

\newcommand{\multimodal}{\textit{multimodal}}

\begin{document}
\rmfamily

\title{Non-reciprocal frequency conversion in a multimode nonlinear system}

\author{Sahil Pontula$^{1,2,3}$}
\email{spontula@mit.edu}
\author{Sachin Vaidya$^{1}$}
\author{Charles~Roques-Carmes$^{3,4}$}
\author{Shiekh Zia Uddin$^{1,3}$}
\author{Marin Solja\v{c}i\'{c}$^{1,3}$}
\author{Yannick Salamin$^{1,3,5}$}
\email{yannick.salamin@ucf.edu}

\affiliation{
$^1$ Department of Physics, MIT, Cambridge, MA 02139, USA. \\
$^2$Department of Electrical Engineering and Computer Science, MIT, Cambridge, MA 02139, USA.\\
$^3$ Research Laboratory of Electronics, MIT, Cambridge, MA 02139,  USA.\\
$^4$ E. L. Ginzton Laboratories, Stanford University, 348 Via Pueblo, Stanford, CA 94305, USA.\\
$^5$ CREOL, The College of Optics and Photonics, University of Central Florida, Orlando, Florida 32816, USA.
}

\begin{abstract}

Nonlinear optics has become the workhorse for countless applications in classical and quantum optics, from optical bistability to single photon pair generation. However, the intrinsic weakness of optical nonlinearity has meant that large input powers and weak output powers are often a necessity in nonlinear frequency conversion. Here, motivated by recent advances in using non-Hermitian photonics and gain/loss engineering to enable non-reciprocal light transport, we explore how the interplay between non-Hermiticity and optical nonlinearity leads to a fundamentally new regime of nonlinear frequency conversion. We show how non-Hermitian coupling between discrete frequency modes can result in non-reciprocal flow of energy in the frequency dimension, closely resembling the non-Hermitian skin effect (NHSE). Applying our theory to a multimode nonlinear cavity supporting cascaded nonlinear processes, we create an asymmetric infrared (IR) comb that features a ``skin'' frequency mode populated with efficiency exceeding 85\%. Furthermore, we demonstrate how three-wave mixing processes in the non-reciprocal infrared comb we generate enables terahertz (THz) generation exceeding the Manley-Rowe limit. We then show how the non-reciprocal frequency conversion is robust against cavity defects and disorder that cause random fluctuations in the dissipation rate for different modes. Moreover, in certain regimes, the nonlinear, non-Hermitian system supports stable limit cycles that can enable multimode pulsing with picosecond pulse widths and GHz repetition rates. Finally, we explore how the system can be applied to generate simultaneous IR and THz frequency combs, potentially unlocking novel applications in spectroscopy and metrology. 


\end{abstract}
\maketitle


\section{\label{sec:level1}Introduction}

Nonlinear optical systems have long been a cornerstone of advanced photonics, enabling a wide range of applications from frequency conversion \cite{fejer1994nonlinear, boyd2008nonlinear} and high-speed communication \cite{schneider2013nonlinear} to quantum information processing \cite{chang2014quantum, caspani2017integrated}. 
More recently, the focus has shifted to multimode nonlinear systems, which offer the potential for more complex interactions through nonlinear processes. For instance, frequency combs, which consist of many discrete frequency modes, have found widespread application in precision metrology and high-capacity data transmission \cite{picque2019frequency, marin2017microresonator, corcoran2020ultra}. However, traditional nonlinear optical systems are constrained by the intrinsic optical properties of nonlinear materials, which are generally weak and obey laws of reciprocity. This restricts conversion efficiencies and the ability to manipulate nonlinear energy flow for more sophisticated functionalities. Achieving control over nonlinear interactions in multimode systems could enable breakthroughs such as overcoming the standard efficiency limits in frequency conversion \cite{salamin2021overcoming} and realizing new topological effects in a synthetic frequency dimension \cite{suh2024photonic}. Despite these promising applications, control over multimode nonlinear frequency conversion has been difficult to achieve owing to high dimensionality and the intrinsic complexity of many nonlinear interactions. 



The introduction of non-Hermitian elements into optical systems has opened new avenues for controlling light-matter interactions and energy flow in real and synthetic dimensions \cite{sounas2017non, caloz2018electromagnetic, yuan2018synthetic,el2019dawn, asadchy2020tutorial, lustig2021topological, wang2023non}. These systems can exhibit unique phenomena not seen in traditional Hermitian systems, such as exceptional points, non-reciprocity, parity-time ($\mathcal{P}\mathcal{T}$) symmetry breaking, and the non-Hermitian skin effect (NHSE) \cite{de2024solitons}. 
Recent studies have shown that breaking reciprocity in a discrete synthetic frequency dimension can realize non-Hermitian Hamiltonians and topological windings in the photonic band structure \cite{wang2021generating, wang2021topological}. 
However, the intersection of non-reciprocity in a frequency dimension and nonlinear frequency conversion has not been explored, although exciting related work in the optomechanical domain has recently been reported \cite{shen2023nonreciprocal}.

In this work, we investigate a multimode nonlinear cavity system that leverages both Hermitian and non-Hermitian interactions to achieve enhanced control over nonlinear energy flow. Our system features a cavity supporting multiple frequency modes that are nonlinearly coupled to a common idler mode. This configuration supports cascaded parametric upconversion and downconversion processes, forming a frequency comb. 
We demonstrate that the interplay between nonlinear Hermitian coupling, dissipation, and anti-Hermitian amplitude modulation can be utilized to shape nonlinear frequency conversion processes. 
By fine-tuning the balance between nonlinearity, dissipation, and amplitude modulation, we achieve non-reciprocal frequency conversion and enhanced energy localization in a ``chiral mode.'' This boosts both the chiral mode conversion efficiency and the idler mode conversion efficiency, as a byproduct of the cascaded nonlinear processes. Our results further reveal how nonlinear frequency conversion can be controlled using system parameters to generate frequency combs of arbitrary asymmetry, with the asymmetry remaining robust to defects and disorder in the frequency comb. We also identify different behaviors (``phases'') in the nonlinear, non-Hermitian system's mean-field dynamics, demonstrating stable multimodal limit cycles that could provide a new resource for multimodal picosecond pulse generation. Finally, we describe how a variant of our system can enable simultaneous flat-top comb generation at IR and THz frequencies, paving the way towards perfect sinc-shaped pulse generation in two disparate frequency regimes from a continuous source.

\section{\label{sec:level2}Theory and system description}

\begin{figure}
    \centering
    \includegraphics[scale=0.8]{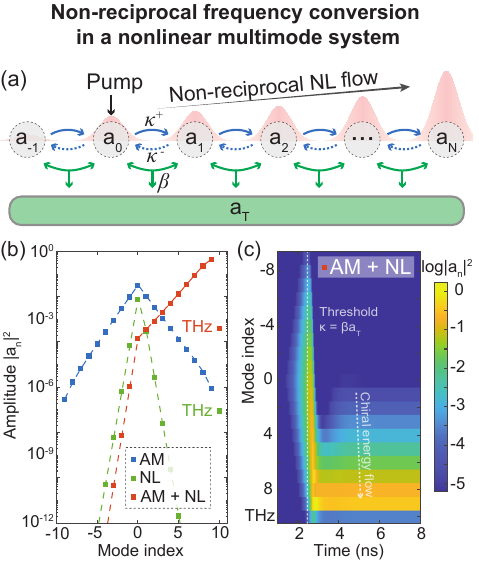}
    \caption{\textbf{Shaping nonlinear energy flow in a non-Hermitian nonlinear multimode cavity.} (a) Tight-binding schematic of nearest-neighbor couplings in a frequency multimode system with second-order nonlinearity. The interplay between nonlinear Hermitian coupling (NL) mediated by a common idler mode (here at THz frequency) and the anti-Hermitian coupling generated by amplitude modulation (AM) at the idler frequency breaks reciprocity in the frequency conversion ($\kappa_+\ne \kappa_-$), which can create unidirectional energy flow in the frequency dimension. (b) The presence of anti-Hermitian coupling through AM or Hermitian coupling through NL alone is insufficient to break symmetry in the frequency conversion process. However, the presence of both coupling simultaneously generates an effective non-Hermitian system that can bias frequency downconversions and create an asymmetric frequency comb. (c) In the nonlinear, non-Hermitian system, the interplay between NL and AM is reflected in the temporal dynamics of the modal amplitudes since the nonlinear coupling is time-dependent, $\beta a_T(t)$. The initial frequency conversion is symmetric since $\kappa_+=- \kappa_-$ (purely AM coupling). As the terahertz mode is populated, $\kappa\rightarrow \beta a_T$. A threshold is reached where reciprocity in upconversion/downconversion is broken and an NHSE-type phenomenon in the frequency dimension occurs, with all energy flowing to the chiral mode $a_N$. In (b)/(c), $N=19$, $\beta=9.40$ s$^{-1}$, $\kappa=7.8\times 10^9$ s$^{-1}$, $Q_0=9\times 10^5$, $\mu_0=0.1\gamma_0$, $Q_T=10^3$, and $|s_0|^2=5$ MW. In this system, the pump frequency $\omega_0=2\pi\cdot 282$ THz, and the idler frequency $\omega_T=2\pi\cdot 1.06$ THz. The modal energies $|a_n|^2$ are given in J.}
    \label{fig:fig1}
\end{figure}

The system under consideration is depicted in Fig. \ref{fig:fig1}a, representing a \multimodal~nonlinear cavity that supports multiple discrete frequency modes, all nonlinearly coupled to a common idler bath mode at frequency $\omega_T$. A pump mode at $\omega_0$ initiates cascaded parametric upconversions and downconversions mediated by the common idler mode to create blueshifted and redshifted modes relative to the pump mode, forming a comb with spacing $\omega_T$. The non-interacting multimode driven-dissipative Hamiltonian describing this system is the sum of the bare Hamiltonians for each mode and reads \cite{gardiner1985input}:
\begin{align}
\begin{split}
        \mathcal{H}_0/\hbar = i\sqrt{2\gamma}s_0(\hat a_0^\dagger-\hat a_0) &+ \sum_{n} [\omega_n-i(\gamma+\mu)] \hat a_n^\dagger \hat a_n\\
        &+ [\omega_T-i(\gamma_T+\mu_T)] \hat a_T^\dagger \hat a_T, 
\end{split}
\end{align}
where $a_n$ is the annihilation operator for the frequency mode at angular frequency $\omega_n$ and $a_T$ is the annihilation operator for the idler bath mode at frequency $\omega_T$. This Hamiltonian is non-Hermitian due to on-site loss ($\gamma$ for outcoupling loss and $\mu$ for intrinsic loss) that emerges from coupling of each frequency mode to the environment. Here, we only consider the mode at $\omega_0$ to be externally pumped ($s_0\ne 0$). The nonlinear interaction between the frequency modes is described by a Hermitian Hamiltonian \cite{graham1968quantum}
\begin{align}
    \mathcal{H}_{\mathrm{NL}}= i\beta\sum_{n} \hat a_T^\dagger  \hat a_n^\dagger \hat a_{n-1} + \mathrm{h.c.},
\end{align}
where $\mathrm{h.c.}$ denotes the Hermitian conjugate. We now add amplitude modulation at the frequency $\omega_\mathrm{mod}=\omega_T$ to this system, generating an anti-Hermitian Hamiltonian \cite{yuan2021synthetic}
\begin{align}
    \mathcal{H}_{\mathrm{A}} = i\kappa\sum_n \hat a_n^\dagger \hat a_{n-1} - \mathrm{h.c.}
\end{align}
The composite Hamiltonian $\mathcal{H}=\mathcal{H}_0+\mathcal{H}_{\mathrm{NL}}+\mathcal{H}_{\mathrm{A}}$ is non-Hermitian and, from it, we can derive equations of motion in the mean field for the annihilation operators $\hat a_n,\hat a_T$, as shown in the S.I.. We have
\begin{align}
    \begin{split}
        \dot a_T = &\beta\sum_k a^*_ka_{k-1} - (\gamma_T + \mu_T) a_T\\
        \dot a_n = &-(\beta^* a_T - \kappa^*) a_{n+1} + (\beta a_T^* + \kappa)a_{n-1} \\
        &- (\gamma + \mu) a_n + \sqrt{2\gamma}s_n\delta_{n,0},
    \end{split}
    \label{eq:nh}
\end{align}
%
which describes the evolution of the mode amplitudes in the multimode system. As we will show, the interplay between on-site non-Hermiticity (loss) and non-Hermitian intermodal coupling in this model will enable us to have a new degree of freedom in controlling nonlinear energy flow in a frequency dimension and shaping nonlinear frequency conversion. 

The nonlinear system of differential equations specified by Eq. \ref{eq:nh} can be numerically solved as a function of time starting from an initial condition where the fields are zero. If the system reaches steady state, the power conversion efficiency in a given mode can be computed as
\begin{align}
    \eta_n = \frac{2\omega_n\gamma_n|a_n|^2}{\sum_k 2\omega_k(\gamma_k + \mu_k)|a_k|^2},
    \label{eq:eff}
\end{align}
where the indices $n,k$ can refer to any mode in the system (including the idler mode). This equation holds because losses must equal drives (amplitude modulation and pumping at $\omega_0$) in steady state. It is also sometimes useful to compare the power generation efficiency to the Manley-Rowe (MR) limit \cite{manley1956some}. The MR limit describes the optimal efficiency in a difference-frequency generation process where every high-energy pump photon produces a lower-energy signal and idler photon. In the case of the idler mode, the MR limit is then given by $\omega_T/\omega_0$.

\section{\label{sec:level3}Chiral energy flow in nonlinear multimode cavities}

\begin{figure*}[t]
    \centering
    \includegraphics[scale=0.8]{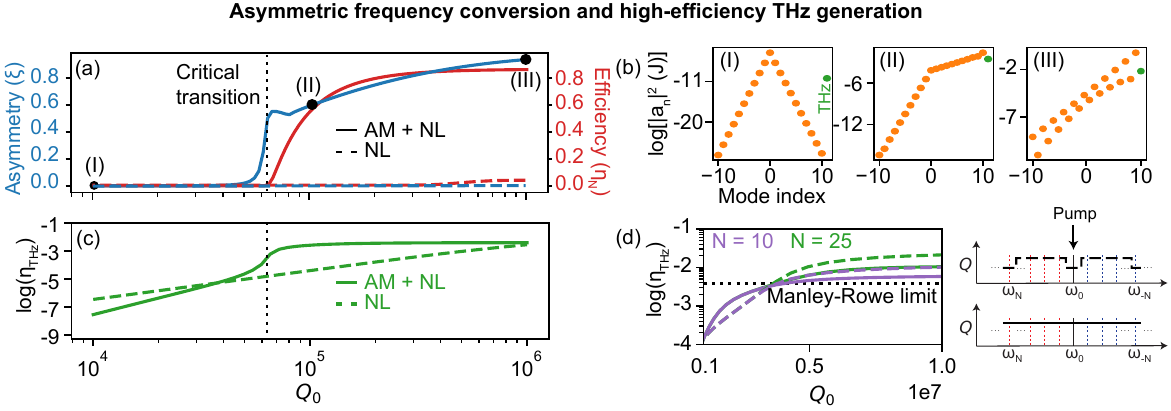}
    \caption{\label{fig:2}\textbf{Asymmetric and highly efficient frequency conversion through non-reciprocal frequency conversion.} (a) By sweeping a parameter of the nonlinear system, one can control the degree of asymmetry $\xi$ in the frequency comb. Here, we sweep the dissipation ($Q$ factor) of the IR modes, changing the interplay between nonlinear (NL) coupling and amplitude modulation (AM), which generates non-Hermitian coupling. When $Q_0$ is low, the nonlinear coupling is very weak and AM creates a symmetric comb, $\xi\approx 0$ (in this case, the on-site loss $\gamma\gg\kappa$ so that modes other than the pump mode at $\omega_0$ are negligibly occupied). As $Q_0$ is increased, the combined effect of AM and NL begins to dominate the system dynamics (over on-site loss). Intuitively, one enters the regime of a high-finesse cavity in frequency space. The combined effect of AM and NL suppresses coupling towards higher frequency, biasing the frequency comb towards lower frequency modes. Finally, for the highest $Q_0$, nearly complete asymmetry is obtained, $\xi\approx 1$. The interference pattern in the modal energy distribution for this case emerges due to non-open boundary conditions at the frequency boundaries $\omega_{\pm N}$, which results in Bloch mode interference in the frequency space lattice. The efficiency parameter $\eta_N$ quantifies how much power is converted into the chiral mode. A sharp increase near the transition to high asymmetry is observed, and the efficiency asymptotes to around 86\% for high $Q_0$. (b) Varying levels of asymmetry in the modal energy distribution for three different values of the $Q_0$ sweep shown in (a). (c) As the asymmetry is increased through improving the finesse of the frequency space cavity, so too is the THz conversion efficiency. The non-reciprocity strongly biases THz generation (difference frequency generation) over THz annihilation (sum frequency generation) processes. (d) The THz conversion efficiency can surpass the Manley-Rowe limit, with highest efficiencies achieved for longer combs (more THz-generating cascading steps, green) and high $Q$, dissipation-engineered frequency-space cavities (dashed line). In simulations (a)-(c), $|s_0|^2=1$ MW, $\beta=4.70$ s$^{-1}$, $\kappa=2\pi\cdot 1060$ MHz, $Q_T=10^4$, and $N=10$. An intrinsic loss $\mu=0.01\gamma$ was assumed. In the absence of AM, a seed ($|s_1|^2=1$ MW) was added to form the frequency comb. In this system, $\omega_0=2\pi\cdot 282$ THz, $\omega_T=2\pi\cdot 1.06$ THz. In (d), $\kappa=2\pi\cdot 10.6$ MHz and for the dissipation-engineered cavity, $Q_{\pm N}=Q_0=10^6$ is fixed.}
    \label{Fig2}
    \label{fig:fig2}
\end{figure*}

A key prediction of non-Hermitian point gap topology is the non-Hermitian skin effect (NHSE), which occurs when the hopping reciprocity in a lattice is broken, enabling bulk eigenstates of the Hamiltonian to be localized on the boundary of the lattice \cite{okuma2020topological, zhang2022review, lin2023topological}. In one dimension, the Hatano-Nelson model is a simple example exhibiting the NHSE, where reciprocity is broken via unequal nearest-neighbor rightward and leftward hoppings~\cite{hatano1996localization, kawabata2020higher}. In our system, we realize a non-linear analogue of the Hatano-Nelson model, where nonlinear Hermitian coupling and non-Hermitian amplitude modulation combine to give non-reciprocal hopping amplitudes in the frequency dimension. In Fig. \ref{fig:fig1}b,c, we show how this can result in unidirectional (or ``chiral'') energy flow towards the lowest-frequency mode in  our ``frequency lattice.'' 

When we pump the mode at $\omega_0$ and with $\kappa\approx 0$ (small amplitude modulation), the result is a symmetric comb around $\omega_0$ with steady state modal energy dropping on either side of $\omega_0$. As $\kappa$ increases, the comb becomes asymmetric, pulling up the redshifted branch of the comb until modal energy eventually increases monotonically with mode index (lower frequency). To quantify the asymmetry in the redshifted and blueshifted branches of the comb around the pump mode $\omega_0$, we will assume a long frequency comb with negligible boundary effects. In this case, we can solve the rate equations in steady state using an ansatz $a_{n+1}\propto r_1 a_n,a_{-n+1}\propto r_2 a_{-n}$ respectively for the redshifted and blueshifted branches ($n>0$ here). This ansatz is physically motivated by the non-Hermitian analog of Bloch bands (where generally $r_1=r_2=e^{ik}$ with $k$ the wavenumber). Both $r_1,r_2$ must satisfy the steady state condition
\begin{equation}
    \left(\kappa-\beta a_T\right)r + \left(\kappa+\beta a_T\right)r^{-1} - \gamma = 0,
\end{equation}
where $a_T$ denotes the steady state THz mode amplitude. From the solutions of this equation, we can then define the asymmetry parameter as
\begin{align}
\begin{split}
     \xi = \frac{\log |r_1| + \log |r_2|}{2\log |r_2|},
\end{split}
\end{align}
as detailed in the S.I.. Graphically, in a logarithmic plot of modal energy, $\xi$ represents a normalized ratio of the sum of the slopes of the redshifted and blueshifted branches. When amplitude modulation is absent ($\kappa=0$) and dissipation exceeds the nonlinear rate ($\gamma\gg \beta a_T$), $|r_1|=1/|r_2|$, generating a symmetric comb with $\xi=0$. When the strength of amplitude modulation approaches the nonlinear rate, $\kappa\sim\beta a_T$, reciprocity is broken in the upconversion and downconversion processes, biasing energy flow towards redshifted modes and creating an asymmetric comb with $\xi\ne 0$ (see S.I. for details). We confirmed the validity of $\xi$ as a figure of merit for asymmetry by numerically performing exponential fits to the modal energy distributions (see S.I. for details).

We now show how the asymmetry in the frequency comb can be controlled through the interplay between nonlinear and non-Hermitian coupling. In Fig. \ref{fig:fig2}a, we plot the asymmetry parameter $\xi$ over a sweep of the quality factor of the frequency comb modes, $Q_0$. When the comb modes are very leaky, the nonlinear coupling remains weak and a symmetric comb is created through the amplitude modulation (Fig. \ref{fig:2}b, (I)). As $Q_0$ is increased, nonlinear energy can more effectively flow in the frequency space lattice. The asymmetry initially increases slowly, then undergoes a sharp increase as the redshifted part of the comb ($n>0$) is pulled upwards. This occurs because $\beta a_T\rightarrow \kappa$, amplifying coupling to lower-frequency modes while simultaneously suppressing the nearest-neighbor hopping to higher-frequency modes. 

When the asymmetry reaches $\xi=0.5$, hopping to higher frequency modes is completely suppressed and a flat comb for the redshifted cascading orders is created through a balance between gain and loss for each mode. The steady state energy of the blueshifted modes continue to decay exponentially away from the pump mode $\omega_0$ as discussed above (Fig. \ref{fig:2}b, (II)). For the highest $Q_0$ considered, nearly perfect asymmetry in the frequency comb is generated (Fig. \ref{fig:2}b, (III)). The interference pattern in the steady state modal energy distribution for this last case (inset) is due to Bloch interference in the frequency lattice that emerges because the boundaries of this system in the frequency dimension are not impedance matched (i.e. are partially reflecting) \cite{hu2022mirror}. The interference exists because we only change $Q_0$ in these simulations, but can be removed by additionally tuning $\beta,\kappa$ \footnote{Therefore, boundary effects become important for the highest $Q_0$ and can invalidate the method for calculating asymmetry described above.}.

Also shown in Fig. \ref{fig:fig2}a is a plot of the power conversion efficiency for the chiral mode $\eta_N$. We observe a sharp increase in $\eta_N$ near the transition to high asymmetry. $\eta_N$ asymptotes near 86\% for high $Q_0$, but we note that higher efficiencies may be obtained by impedance matching the frequency comb to eliminate interference in the modal energy distribution. 

\subsection{High power terahertz generation}

\begin{figure}[t]
    \centering
    \includegraphics[scale=0.8]{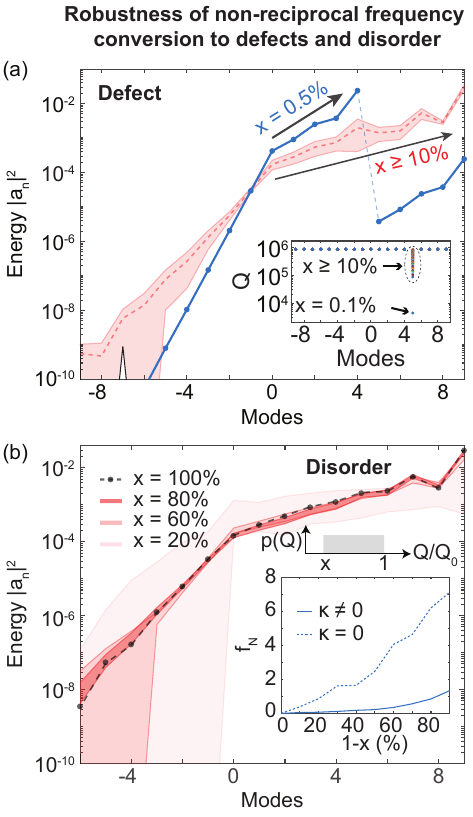}
        \caption{\textbf{Topological robustness of non-reciprocal frequency conversion to defect and disorder.} (a) A random low $Q$ factor defect is introduced in the frequency comb. The non-reciprocal frequency conversion is robust to strong defects, though the strongest defects result in the skin effect terminating just before the defect mode (which acts as a ``frequency mirror'' \cite{hu2022mirror}). Here, the modal energy $|a_n|^2$ is plotted for various defect strengths $Q_{\mathrm{defect}}/Q_0=x$, where $Q_0$ denotes the $Q$ factor of the remaining modes in the frequency comb. (b) Robustness of non-reciprocal frequency conversion to disorder. Here, the $Q$ factor of all frequency modes in the comb is sampled uniformly from $[xQ_0,Q_0]$. $N_\mathrm{samp}=100$ samples are drawn from this uniform distribution and individually simulated, with the shaded regions denoting $\pm 1\sigma$ standard deviation from the mean modal energies. Shown in the inset is the robustness of the energy of the chiral mode $a_N$, plotted as the standard deviation normalized by the mean (coefficient of variation), $f_N\equiv \sigma_N /\langle |a_N|^2\rangle$, where $\langle |a_N|^2\rangle$ denotes the ensemble average over $N_\mathrm{samp}$ samples. In the absence of amplitude modulation $\kappa=0$, no skin effect is present and the disorder affects $f_N$ much more strongly than the case $\kappa\ne 0$, where the skin effect helps stabilize the modal energy in the chiral mode.}
    \label{fig:fig3}
\end{figure}

The unidirectional energy flow enabled by non-reciprocal frequency conversion in the frequency dimension biases frequency downconversions that produce THz photons (and suppresses frequency upconversions that annihilate THz photons) in the cascaded nonlinear system. We predict this to lead to high energy in the THz idler mode of the multimode system. 

This concept is shown in Fig. \ref{fig:fig2}c, where the efficiency of conversion into the propagating THz idler mode outcoupled from the cavity increases with $Q_0$. Notice that the curve of THz efficiency plateaus near $\xi\approx 0.5$. At this value of the asymmetry parameter, upconversions are perfectly suppressed, a flat IR comb is generated, and THz generation is enhanced by the number of comb modes. Here, the THz photon generation rate is equally contributed by three-wave mixing between each pair of nearest-neighbor comb modes. As $Q_0,\xi$ increase, the THz generation becomes dominated by three-wave mixing at the low frequency end of the frequency comb, which is populated strongly due to the non-reciprocal frequency conversion. 


We observe an enhancement in THz conversion efficiency for the nonreciprocal frequency comb compared to the reciprocal frequency comb without amplitude modulation. (The latter system requires a seed to generate comb modes and the THz idler mode, so we set $|s_1|^2=|s_0|^2$ for this system only.) We attribute this enhancement to the unique ability to suppress upconversions in the non-Hermitian system and therefore break symmetry between THz-generating and THz-annihilating processes. The critical transition to high asymmetry and chiral mode efficiency is accompanied by a transition (black dotted line in Fig. \ref{fig:2}c) to high THz conversion efficiencies. 

In Fig. \ref{fig:2}d, we show how THz conversion efficiencies surpassing the Manley-Rowe limit can be achieved through (1) longer combs and (2) appropriate dissipation engineering. The former enables more cascading steps to generate THz photons from a single pump photon, while the latter shapes the boundary conditions in the frequency lattice to maximize THz-generating nonlinear interactions. In Fig. \ref{fig:2}d, the ``dissipation engineering'' is not optimized and merely corresponds to a realistic system where the frequency lattice is terminated by modes of lower $Q$ factor (the pump mode also has lower $Q$ factor due to larger incoupling/outcoupling loss).


We highlight that the high power THz generation in the nonlinear, non-Hermitian system is enabled through the interplay between optical nonlinearity and non-Hermitian coupling between frequency modes. Although non-reciprocal frequency conversion can be achieved through combined amplitude and phase modulation without optical nonlinearity (as described in 
the S.I.), the latter is necessary to generate THz idler photons. 


\section{\label{sec:level4} Robustness to defect and disorder}

In this section, we show how the non-reciprocal frequency conversion we observe in the frequency dimension preserves the robustness to disorder observed for NHSE \cite{claes2021skin}. In Fig. \ref{fig:fig3}a, we introduce a random leaky mode in the frequency mode which has a $Q$ factor lower than the other modes in the comb, $Q_\mathrm{defect}<Q_0$. The original skin effect is preserved for large defects strengths, even when the defect mode has a $Q$ factor over 90\% below $Q_0$. When the defect mode becomes too leaky ($Q_\mathrm{defect}/Q_0\lesssim 1\%$), the comb terminates at the defect mode and the new chiral mode becomes the comb mode just before the defect mode (which now acts as a frequency mirror \cite{hu2022mirror}).

In Fig. \ref{fig:fig3}b, we consider disorder in the $Q$ factor distribution for all of the frequency modes in the comb. To model this disorder, we sample each mode's $Q$ factor individually and uniformly from the distribution $[xQ_0,Q_0]$ with $0\le x\le 100\%$. We see how the skin effect is resistant to disorder, preserving the unidirectional flow of energy in frequency space even for large amounts of disorder $x>50\%$. To show that this emerges from the non-Hermitian coupling in the system, we plot in the inset of Fig. \ref{fig:fig3}b the standard deviation in the modal energy of the chiral mode normalized to its mean in the presence ($\kappa\ne 0$) and absence ($\kappa=0$) of amplitude modulation. (The mean and standard deviation are computed over 100 simulated systems with $Q$ factors drawn from the aforementioned uniform $Q$ factor distribution.) The variation in the chiral mode's energy is larger and scales more sharply with the dimensionless disorder parameter $x$ in the case $\kappa=0$, showing the robustness of the non-reciprocal frequency conversion ($\kappa\ne 0$) to the disorder considered here.

\begin{figure}[t]
    \centering
    \includegraphics[scale=0.8]{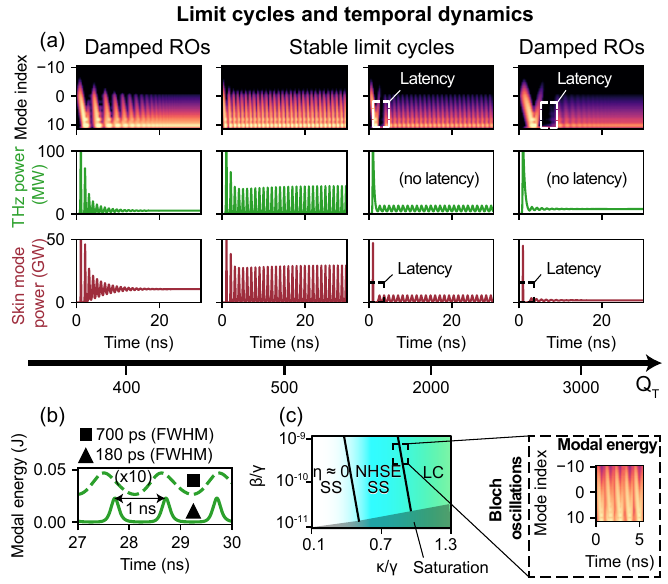}
    \caption{\textbf{Limit cycles and temporal dynamics in the nonlinear, non-Hermitian multimode system.} (a) For low quality factors of the idler bath mode ($Q_T$), the system initially features damped relaxation oscillations (ROs) and the amplitude modulation drives the system towards a stable steady state that features non-reciprocal frequency conversion. As $Q_T$ increases, the system eventually transitions into a regime of stable limit cycles that feature periodic oscillations in modal energy. Finally, larger values of $Q_T$ cause a return back to damped ROs, with the nonlinearity and amplitude modulation enabling a gain-loss equilibrium. Notice in this last case the latency period between the initial pulse and the subsequent ROs in the chiral mode power (this latency is absent for the THz mode). (b) Oscillations in THz modal energy in the limit cycle regime. As $Q_T$ increases, the limit cycles transition from pulses to sinusoidal oscillations, effectively increasing the pulse width. The repetition rate of the limit cycles is around 1 GHz, roughly corresponding to the amplitude modulation strength $\kappa$. (c) Phase diagram showing the steady state (SS) and limit cycle (LC) regimes as the nonlinear and amplitude modulation strengths are swept. For small $\kappa$, a symmetric comb in steady state is produced ($\xi,\eta\approx 0$). For larger $\kappa$, an asymmetric comb in steady state is produced by the non-reciprocal frequency conversion ($\xi,\eta>0$). For the largest $\kappa$, LCs are present. Of additional note is the saturation region for low $\beta$, where the nonlinearity is insufficient to prevent amplitude modulation from driving optical powers to saturation. Along the transition boundary, the temporal dynamics feature critically damped Bloch oscillations where energy bounces back and forth in the frequency space cavity bounded by the terminal comb modes. In (a), (b), parameters used are $Q_0=1.1\times 10^5,\kappa/\gamma=0.856,\beta/\gamma=5.83\times 10^{-10}$. The pump power is 5 MW and 10 redshifted as well as 10 blueshifted modes are simulated. In (c), the pump power is 1 W and $Q_T=500$.}
    \label{fig:fig4}
\end{figure}

\section{\label{sec:level5}Temporal dynamics and stable limit cycles}

In this section, we take a closer look at the temporal dynamics of states supported by the nonlinear, non-Hermitian multimode system considered in the previous sections. In Fig. \ref{fig:fig4}, we show how these states can be broadly classified as steady state solutions or limit cycles (LCs). Thus far, we have focused on combs generated in steady state. LCs are a particularly interesting class of solutions unique to nonlinear systems that are characterized by stable oscillations around a fixed point \cite{strogatz2018nonlinear}. As shown in Fig. \ref{fig:fig4}a, the LC regime can be entered by gradually increasing the quality factor of the THz idler mode, $Q_T$. The steady state solutions are distinguished by an approach to equilibrium characterized by damped relaxation oscillations (ROs), whereas the limit cycle regime is characterized by periodic oscillations in the modal amplitudes that never relax to a steady state. We numerically verified the existence of a fixed point in between the extrema of the LCs. Zooming in on the LC regime, we observe that the repetition rate of the LCs appears to be governed by the amplitude modulation strength $\kappa$ while the pulse width is governed by the interplay between on-site loss and gain/loss due to intermodal coupling. Near the start of the LC regime when sweeping $Q_T$, the LCs resemble picosecond pulses with GHz repetition rates, potentially providing a new regime of frequency-tunable pulsed sources.

We can sweep over the nonlinear and amplitude modulation strengths $\beta,\kappa$ to create a phase diagram portraying the distinct temporal dynamics, as shown in Fig. \ref{fig:fig4}c. For small $\kappa$, symmetric combs are produced in steady state irrespective of $\beta$. As $\kappa$ is increased, the non-reciprocal frequency conversion drives the system towards an asymmetric comb in steady state. The nonlinearity acts as a loss mechanism to balance the gain from amplitude modulation, so that for small nonlinearity $\beta$, gain induced by $\kappa$ drives the system to very high powers that would be limited by saturation effects/damage thresholds. Finally, the largest values of $\kappa$ yield LCs. Like the sharp transition to non-reciprocal frequency conversion, the transition to LCs appears to be quite sensitive to $\kappa$, switching sharply from steady state solutions to LCs for almost all values of $\beta$. Therefore, beyond a certain threshold, the amplitude modulation stabilizes the LCs and the LC phenomenon is relatively insensitive to the nonlinear strength. The phase transition boundary is characterized by critically damped Bloch oscillations in the frequency space cavity defined by the frequency comb, as shown in the inset \cite{englebert2021bloch}. These oscillations are characterized by nonlinear energy flow bouncing back and forth inside the frequency space cavity (outside of the LC regime, these oscillations are damped and relax to a steady state).

\section{IR and terahertz subcomb generation by nonlinear non-Hermitian frequency conversion}
\label{sec:subcomb}

\begin{figure}[t]
    \centering
    \includegraphics[scale=0.8]{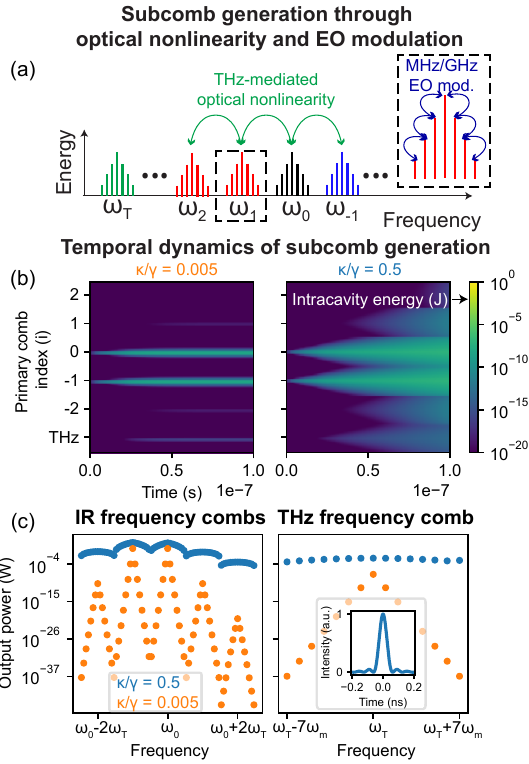}
    \caption{\textbf{Subcomb generation through combined electro-optic (EO) and cascaded $\chi^{(2)}$ nonlinearity.} (a) Through cascaded difference frequency generation initiated by a pump at $\omega_0$ and seed at $\omega_1=\omega_0-\omega_T$, a ``primary'' frequency comb with spacing $\omega_T$ is created. Further electro-optic modulation of this system at a MHz-GHz frequency $\omega_\mathrm{mod}$ creates a ``secondary'' comb (subcomb) at each comb line of the primary comb. The result is the change from generation of a single THz idler mode to a comb of THz idler modes. (b), (c) When the strength of the amplitude modulation $\kappa$ is weak, the combs are negligibly populated in steady state and only a single THz idler mode is generated with weak efficiency. When $\kappa$ is comparable to the decay rates of the IR modes, the primary and secondary combs are sustained, generating a comb of THz idler modes in the steady state with mW-level powers. In these simulations, the system was pumped with $|s_{0,0}|^2=1$ W at $\lambda_0=1064$ nm and seeded with $|s_{1,0}|^2=1$ W at $\omega_1=1068$ nm. The nonlinear parameter was $\beta=4.70$ s$^{-1}$ and $Q$ factors for the IR and THz modes were $Q_n=4\times 10^6,Q_T=5\times 10^3$. The primary and secondary comb spacings were $\omega_T=2\pi\cdot 1.06$ THz and $\omega_\mathrm{mod}=2\pi\cdot 1.00$ GHz. We simulate $I=5$ subcombs, each comprised of $J=15$ modes.}
    \label{fig:subcomb}
\end{figure}

Thus far, in the cascaded nonlinear system, we have considered a single THz idler mode that is populated through nonlinear coupling between nearest-neighbor comb modes in a single comb that is sustained through cascaded second-order nonlinear processes and amplitude modulation at a frequency that coincides with the comb spacing and the THz idler mode frequency, $\omega_\mathrm{mod}=\omega_T$. Because of the current difficulty in accessing THz modulators, we consider in this section phenomena that emerge when modulation is performed instead at MHz-GHz frequencies, so that $\omega_\mathrm{mod}\ll \omega_T$. We will show that this new system is capable of generating terahertz frequency combs with high power. As shown in Fig. \ref{fig:subcomb}a, we begin with a system that employs cascaded difference frequency generation processes initiated by a pump and seed at frequencies $\omega_{0,1}$ (respectively) to create a THz-spaced frequency comb \cite{ravi2016cascaded, hemmer2018cascaded, salamin2021overcoming, pontula2024multimode}. We will term this comb the ``primary'' comb.  By design, phase matching constraints enforce only population of the first-order THz idler mode at $\omega_T$ (so that modes at $2\omega_T,3\omega_T,...$ are not created). Now, we add electro-optic (EO) modulation to the system, turning each line of the primary comb into a ``subcomb'' or ``secondary'' comb. This takes the single THz idler mode generated through difference frequency generation in the primary comb and turns it into a comb of THz idler modes with spacing given by the MHz-GHz EO modulation frequency. Using a tight-binding, coupled-mode analysis, the equations of motion for this system can be derived as
\begin{align}
    \begin{split}
        \dot a^T_{-k} &= \beta\sum_{ij} a^*_{i,j+k}a_{i-1,j} - \gamma_T a^T_k \\
        \dot a_{i,j} &= \left(\sum_k -\beta^* a^T_{-k} a_{i+1,j+k} + \beta a^{T*}_{-k} a_{i-1,j-k} \right) \\
        &+ (\kappa^* - C^*) a_{i,j+1} + (\kappa + C)a_{i,j-1} - \gamma a_{i,j} + \sqrt{2\gamma}s_{i,j}.
    \end{split}
    \label{eq:subcomb}
\end{align}
where $a^T_k$ denotes the modal amplitude of the $k^\mathrm{th}$ mode in the THz comb, and $a_{i,j}$ denotes the modal amplitude of the $j^\mathrm{th}$ mode in the secondary comb generated around the $i^\mathrm{th}$ mode in the primary comb. The linear couplings $\kappa,C$ denote amplitude and phase modulation strengths (respectively) in each IR secondary comb (subcomb) and $\beta$ denotes the nonlinear coupling strength in the primary comb. As before, $s_{i,j}$ accounts for any external pumps to the system. The intrinsic loss $\mu$ is dropped for simplicity here but can be readily incorporated in the coupled-mode equations, as done in Eq. \ref{eq:nh}.

In Fig. \ref{fig:subcomb}b,c, we show the results of numerically simulating an example system where we apply a continuous wave pump and seed at frequencies $\omega_{0,1}$ such that only $s_{0,0},s_{1,0}$ are nonzero. The pump and seed initiate cascading difference frequency generation processes that create the primary frequency comb and populate a single THz idler mode at $\omega_T$. We now turn on an amplitude modulation of strength $\kappa$ (for simplicity, $C=0$ here so that phase modulation is absent). When $\kappa$ is weak, secondary combs are generated with very weak amplitude and only a single THz idler mode is generated with low power since the primary comb is also not strongly populated away from $\omega_{0,1}$. However, when $\kappa$ is strong, the primary and secondary combs approach a flat-top profile, generating a flat-top comb of THz idler modes at high (mW) powers. Realistically, the width of this comb would be limited by phase matching/dispersion constraints, but the $J=15$ modes in the comb could readily be scaled up to hundreds of modes (with dispersion compensation techniques if necessary). We note that flat-top combs have exciting potential for highly efficient data transmission among other applications \cite{soto2013optical}. The simultaneous generation of multiple flat combs at IR and THz frequencies in a single system could therefore be a novel resource for efficient optical computing and communication.

The modulation strength $\kappa$ necessary to create the THz comb can be crudely estimated by treating the nonlinearity as a perturbation, which is generally valid given $\beta a_k^T \ll \kappa$ in the cases considered in this section. Then, to zeroth order, in steady state (assuming $\kappa\in\mathbb{R}$),
\begin{align}
    0=\kappa(a_{i,j+1} + a_{i,j-1}) - \gamma a_{i,j}.
\end{align}
To generate a flat-top comb, we would like $|a_{i,j}|^2\approx |a_{i,j\pm 1}|^2$, so that $\kappa\approx \gamma/2$.

The parameters considered here were chosen to correspond closely with a multimode cavity architecture proposed for generating the primary comb through cascaded nonlinearity \cite{salamin2021overcoming, pontula2024multimode}. The amplitude modulation strength was chosen to satisfy $\kappa/\mathrm{FSR}=\kappa/\omega_\mathrm{mod}\ll 1$ (where FSR denotes the free spectral range of the cavity which is equal to the modulation frequency $\omega_\mathrm{mod}$ here) to correspond to a Mach-Zender interferometer-based amplitude modulation \cite{wang2021generating}. Together with previously proposed techniques of dissipation engineering the multimode cavity to enhance nonlinear rates and THz generation \cite{salamin2021overcoming, pontula2024multimode}, we anticipate even higher powers and more efficient THz comb generation. 

\section{Discussion and outlook}

In this work, we theoretically demonstrated that non-Hermitian, nonlinear photonic systems can be harnessed to control energy flow in frequency space in novel ways. This enables the generation of non-reciprocal frequency combs, high-power generation of tunable frequency modes that are difficult to produce (or are only weakly produced) by other means, and simultaneous generation of frequency combs at vastly different frequency ranges, such as the infrared and terahertz. Furthermore, we have shown how the non-reciprocal frequency conversion we observe is robust to defects and disorder, reminiscent of the non-Hermitian skin effect and inviting the application of non-reciprocal frequency conversion in realistic experimental platforms for high power nonlinear frequency conversion. In particular, we highlight that the addition of amplitude modulation (AM) provides intrinsic gain in the system and therefore can alleviate the necessary pump powers, quality factors, and nonlinear strengths required for conventional nonlinear frequency conversion. 

The interplay between nonlinearity and non-Hermitian coupling in the frequency comb systems studied here offers numerous exciting avenues of research for both fundamental science and photonic applications. For example, the interaction between nonlinearity and non-Hermiticity in the quantum optical regime of nonlinear frequency conversion remains to be explored and may enable the creation of various multimode quantum and non-Gaussian states of light. In addition, tailoring non-reciprocal hopping in a frequency dimension offers an exiting resource for quantum walk studies. Furthermore, the multimode limit cycles explored in this work suggest the possibility of harnessing multistability for switching, entangled state generation, and more. In the classical domain, we envision that combining non-reciprocal conversion with techniques such as dissipation engineering \cite{salamin2021overcoming, pontula2024multimode} will lead to even more precise control over nonlinear energy flow in multimode systems. 

The subcomb example studied in Sec. \ref{sec:subcomb} offers exciting potential for studying topological physics in multiple synthetic dimensions. The subcomb considered here creates two synthetic frequency dimensions (that are coupled through a terahertz comb) and could realize topological states such as surface chiral modes. Furthermore, the synthetic frequency dimensions can be coupled to other real-space (e.g., coupled ring resonators in a real-space chain or lattice) or synthetic (e.g., orbital angular momentum) dimensions to create topological states of even higher order, such as photonic high-order topological insulators \cite{noh2018topological, xie2018second, dutt2020higher, kim2020recent, li2020higher, kirsch2021nonlinear, schulz2022photonic}.

We briefly comment on experimental platforms capable of realizing the effects described here. Ring resonators have recently received significant attention in studies of topological effects in synthetic frequency dimensions \cite{dutt2020single, wang2021generating, lustig2024emerging}. In particular, these systems used Mach-Zender-based amplitude modulation of a fiber ring resonator to realize non-Hermitian band topology. We view this platform as an excellent candidate to realize the effects described here when combined with lithium niobate photonics (or another nonlinear material) to generate nonlinear coupling. On-chip nonlinear microring resonators have also been extensively developed and suggest exciting potential for developing integrated non-reciprocal frequency conversion setups \cite{chen201312, guarino2007electro, zhang2019broadband}. Finally, advances in terahertz modulation using 2D materials also suggest exciting possibilities for using terahertz amplitude modulators to generate the terahertz-spaced non-reciprocal combs considered in this work \cite{ma2019modulators, wang2021recent}.


\section{Acknowledgements} S.P. acknowledges the financial support of the Hertz Fellowship Program and NSF Graduate Research Fellowship Program. Y.S. acknowledges support from the Swiss National Science Foundation (SNSF) through the Early Postdoc Mobility Fellowship No. P2EZP2188091. S.V. and M.S.\ acknowledge support from the U.S.\ Office of Naval Research (ONR) Multidisciplinary University Research Initiative (MURI) under Grant No.\ N00014-20-1-2325 on Robust Photonic Materials with Higher-Order Topological Protection. C.R.-C. is supported by a Stanford Science Fellowship. This material is based upon work supported in part by the Air Force Office of Scientific Research under the award number FA9550-20-1-0115; the work is also supported in part by the U. S. Army Research Office through the Institute for Soldier Nanotechnologies at MIT, under Collaborative Agreement Number W911NF-23-2-0121. This work is additionally supported in part by the DARPA Agreement No. HO0011249049. We also acknowledge support of Parviz Tayebati.

\bibliography{bib_main}

\end{document}